\documentstyle[12pt]{article}
\begin{document}
\title{AVERAGE ENTROPY OF A SUBSYSTEM\thanks
{Alberta-Thy-22-93, gr-qc/9305007, Phys.\ Rev.\ Lett. {\bf 71}, 1291
(1993).}}
\author{ Don N. Page\\
CIAR Cosmology Program\\
Theoretical Physics Institute\\
Department of Physics\\University of Alberta\\
Edmonton, Alberta\\Canada T6G 2J1\\
Internet:  don@page.phys.ualberta.ca}
\date{(1993 May 7, minor revisions Aug. 5)}
\maketitle
\large
\begin{abstract}
\baselineskip 25pt
If a quantum system of Hilbert space dimension $mn$ is in a random
pure state,
the average entropy of a subsystem of dimension $m\leq n$ is
conjectured to be
$S_{m,n}=\sum_{k=n+1}^{mn}\frac{1}{k} - \frac{m-1}{2n}$ and is shown
to be
$\simeq \ln m - \frac{m}{2n}$ for $1\ll m\leq n$.  Thus there is less
than
one-half unit of information, on average, in the smaller subsystem of
a total
system in a random pure state.
\\
\\
PACS numbers: 05.30.Ch, 03.65.-w, 05.90.+m
\end{abstract}
\normalsize
\pagebreak
\baselineskip 24.2pt

     One natural way to get entropy, even for a system in a pure
quantum state,
is to make the coarse graining of dividing the system into subsystems
and
ignoring their correlations.  For example, suppose a system $AB$ with
Hilbert
space dimension $mn$ and normalized density matrix $\rho$ (a pure
state $\rho =
|\psi\rangle\langle\psi|$ if $\rho^2 = \rho$, which is equivalent to
$Tr \rho^2
= 1$) is divided into two subsystems, $A$ and $B$, of dimension $m$
and $n$
respectively.  (Without loss of generality, take the first subsystem,
$A$, to
be the one with the not larger dimension, so $m\leq n$.)  The density
matrix of
each subsystem is obtained by the partial trace of the full density
matrix
$\rho$ over the other subsystem, so
	\begin{equation}
        \rho_{A}={tr}_B\rho,
	\end{equation}
	\begin{equation}
	\rho_{B}={tr}_A\rho.
	\end{equation}
The entropy of each subsystem is
	\begin{equation}
	S_A=-tr\rho_A \ln \rho_A,\;\;S_B=-tr\rho_B \ln \rho_B.
	\end{equation}
Unless the two systems are uncorrelated in the quantum sense (which
corresponds
to the case that $\rho=\rho_A\otimes \rho_B$), the sum of the
entropies of the
subsystems, which is a coarse graining that ignores the correlations,
is
greater than the fine-grained entropy $S_{AB}$ of the total system:
	\begin{equation}
	S_A+S_B > S_{AB} \equiv -tr\rho \ln \rho.
	\end{equation}
In fact, the three entropies $S_A$, $S_B$, and $S_{AB}$ obey the
triangle
inequality \cite{A}, so if the entire system $AB$ is in a pure state,
which has
$S_{AB}=0$, then $S_A=S_B$, which is an immediate consequence of the
well-known
fact that $\rho_A$ and $\rho_B$ then have the same set of nonzero
eigenvalues.

It may be of interest to calculate how much entropy one typically
gets by this
coarse graining \cite{L1,LP,Pr,P}.  For example, what is the average,
which I
shall call
	\begin{equation}
	S_{m,n} \equiv \langle S_A\rangle,
	\end{equation}
of the entropy $S_A$ over all pure states $\rho =
|\psi\rangle\langle\psi|$ of
the total system?  Here the average is defined with respect to the
unitarily
invariant Haar measure on the space of unit vectors $|\psi\rangle$ in
the
$mn$-dimensional Hilbert space of the total system, which is
proportional to
the standard geometric hypersurface volume on the unit sphere
$S^{2mn-1}$ which
those unit vectors give when the $mn$-complex-dimensional Hilbert
space is
viewed as the $2mn$-real-dimensional Euclidean space \cite{L1}.  Note
that
since $S_A$ is a nonlinear function of the density matrix $\rho_A$,
the average
$\langle S_A\rangle$ of this entropy function is {\it not} the same
as this
function evaluated for the average density matrix
$\langle\rho_A\rangle = {\bf
I}/m$ (the identity matrix acting on the subsystem Hilbert space,
divided by
its dimension $m$), which would be an entropy of $S_{max} \equiv\ln
m$, the
maximum entropy the subsystem $A$ can have.  It is convenient to
define the
average information of the subsystem as the deficit of the average
entropy from
the maximum,
	\begin{equation}
	I_{m,n} \equiv S_{max} - \langle S_A\rangle = \ln m -
S_{m,n}.
	\end{equation}

Lubkin \cite{L1} calculated that, in my notation,
	\begin{equation}
	\langle tr\rho_A^2\rangle = \frac{m+n}{mn+1},
	\end{equation}
and hence he estimated that for $m\ll n$,
	\begin{equation}
	S_{m,n} \simeq \ln m - \frac{m^2-1}{2mn+2}.
	\end{equation}
However, he was unable to calculate $S_{m,n}$ exactly.  Lloyd and
Pagels
\cite{LP}, apparently unaware of Lubkin's work, as I was also when I
did my
calculations, made progress from a slightly different angle by
calculating the
probability distribution of the eigenvalues of $\rho_A$ for random
pure states
$\rho$ of the entire system.  The result, after inserting the
differentials
that were used in the calculation \cite{Lpri} but which were not
given
explicitly in the paper \cite{LP}, and after changing variables to
the
eigenvalues $p_i$, is
	\begin{equation}
	P(p_1,\ldots, p_m)dp_1 \ldots dp_m \propto
	\delta(1-\sum_{i=1}^{m}p_i)
	\prod_{1\leq i<j\leq m} (p_i-p_j)^2
	\prod_{k=1}^{m} (p_k^{n-m}dp_k).
	\end{equation}
The normalization constant for this probability distribution is given
only
implicitly by the requirement that the total probability integrate to
unity.
Although they also did not calculate $S_{m,n}$ exactly, Lloyd and
Pagels
\cite{LP} came to the same qualitative conclusion as Lubkin
\cite{L1}, that for
$m\ll n$ the typical entropy of the (much) smaller subsystem is very
nearly
maximal.

Here I shall show that for $1\ll m\leq n$,
	\begin{equation}
	S_{m,n} \simeq \ln m - \frac{m}{2n},
	\end{equation}
which agrees with Eq. (8) above from Lubkin \cite{L1} in the region
of overlap
but does not require $m\ll n$.  An exact calculation for $m=2$ and
$m=3$ led me
to the tentative conjecture that the exact general formula for $m\leq
n$ is
	\begin{equation}
	S_{m,n}=\sum_{k=n+1}^{mn}\frac{1}{k} - \frac{m-1}{2n},
	\end{equation}
which rather remarkably agrees with what I later calculated exactly
for $m=4$
and $m=5$, so I now think it would be surprising if it were not
always correct,
though I have not yet found a proof for this conjecture.
Furthermore, for
large $n$, the asymptotic expansion for Eq. (11) is
	\begin{equation}
	S_{m,n}=\ln m - \frac{m^2-1}{2mn}
	+ \sum_{j=1}^{\infty} B_{2j} \frac{m^{2j}-1}{2jm^{2j}n^{2j}},
	\end{equation}
where $B_{2j}$ are the Bernoulli numbers, which fits both Eqs. (8)
and (10)
when $n\gg 1$.

To calculate $S_{m,n}$, it is convenient to define $q_i = rp_i$ and
	\begin{eqnarray}
	Q(q_1,\ldots,q_m)dq_1\ldots dq_m &\equiv &
	\prod_{1\leq i<j\leq m} (q_i-q_j)^2
	\prod_{i=1}^{m} (e^{-q_i}q_i^{n-m}dq_i) \nonumber \\
	 &\propto &e^{-r}r^{mn-1}P(p_1,\ldots,p_m)dp_1\ldots
dp_{m-1}dr.
	\end{eqnarray}
Then
	\begin{eqnarray}
	S_{m,n} \equiv \langle S_A\rangle &=&
	-\int (\sum_{i=1}^{m} p_i \ln p_i)P(p_1,\ldots,p_m)dp_1\ldots
p_{m-1}
	\nonumber \\&=&
	\psi(mn+1)-\frac{\int (\sum_{i=1}^{m}q_i \ln q_i)Q dq_1\ldots
dq_m}
	{mn\int Q dq_1\ldots dq_m},
	\end{eqnarray}
using, for integral $N$,
	\begin{equation}
	\int_{0}^{\infty}e^{-x}x^N\ln x dx = N!\psi (N+1) =
	\Gamma'(N+1) = N!(\sum_{k=1}^{N}\frac{1}{k} - \bf C),
	\end{equation}
where $\bf C$ is Euler's constant, which, after some algebra, one can
see
cancels out from the final expression for $S_{m,n}$, leaving a
rational number
for each pair of integers $m$ and $n$.

One can readily calculate by hand that for $n\geq 2$,
	\begin{equation}
	S_{2,n}=\sum_{k=n+1}^{2n-1}\frac{1}{k},
	\end{equation}
which, for example, gives $S_{2,2}=1/3$, slightly less than one-half
$S_{max}=\ln 2$ in that case, and for $n\geq 3$,
	\begin{equation}
	S_{3,n}=\sum_{k=n+1}^{3n}\frac{1}{k} - \frac{1}{n},
	\end{equation}
both of which are fit by Eq. (11), which they suggested as a
generalization.
For $m>3$, the expression for Q is too cumbersome for Eq. (14) to be
readily
evaluated by hand, but I was able to calculate it for $m=4$ and $m=5$
with the
aid of {\it Mathematica} 2.0, after putting in by hand the correct
value of the
integral of Eq. (15), which {\it Mathematica} 2.0 evaluates
incorrectly.  Both
of these values of $m$ also fit Eq. (11).  For $m>5$ I ran into
another
apparent bug in {\it Mathematica} 2.0 which I have not yet figured
out how to
circumvent, but the likelihood that the agreement of my calculations
of
$S_{4,n}$ and $S_{5,n}$ with Eq. (11) is due to accident or error
seems less
than the likelihood that my conjectured Eq. (11) is in fact exact for
all $m$.

However, because I have not yet found any proof of Eq. (11), it
appears
worthwhile to derive an approximate expression for $S_{m,n}$ for
large $m$ and
$n$, which I now do.  In this limit,
	\begin{equation}
	S_{m,n}\simeq -\sum_{i=1}^{m}p_i \ln p_i =
	-\sum_{i=1}^{m}\frac{q_i}{r}\ln \frac{q_i}{r}
	\end{equation}
for $p_i$'s which maximize $P(p_1\ldots p_m)$ or $q_i$'s which
maximize
$Q(q_1,\ldots,q_m)$.  Now
	\begin{equation}
	-\ln Q(q_1,\ldots,q_m)=-\sum_{1\leq i<j\leq m} \ln
(q_i-q_j)^2+
	\sum_{i=1}^{m}[ q_i -(n-m)\ln q_i]
	\end{equation}
is the potential energy for $m$ unit charges on the $q$-line in two
dimensions
due to their mutual electrostatic repulsion, a uniform external unit
electric
field in the negative $q$ direction, and another superposed external
electric
field of strength $(n-m)/q$ in the positive $q$ direction from $n-m$
external
charges fixed at the origin.

For large $m$ and $n$, we can also make the continuum approximation
of
$m\sigma(x)dx$ ``charges" (eigenvalues of the ``most typical" density
matrix
$\rho_A$) in the range $dx$ of the rescaled variable $x=q/m$, so
$\sigma$ is a
normalized linear density (with respect to $x$) of eigenvalues.  The
equilibrium condition (maximization of $Q$ or minimization of the
electrostatic
energy $-\ln Q$) gives the integral equation, with $w\equiv (n-m)/m$,
	\begin{equation}
        \int \frac{\sigma(x)dx}{x-x'} = \frac{w-x'}{x'}
	\end{equation}
for $x'$ in the range where $\sigma(x')>0$, say $0\leq a<x'<b$ for
some
constants $a$ and $b$ that would depend on $w$.  This is a Fredholm
equation of
the first kind \cite{MW}, but with the troublesome feature that the
right hand
side is only given for part of the real axis, $a<x'<b$, where
$\sigma(x')>0$.

The solution can be found  as a special case of the Riemann-Hilbert
problem:
``to find a function, harmonic in a certain plane region $D$,
assuming that on
some parts of its contour we are given the values of the required
function, and
on others the values of its normal derivative" \cite{Mik}.  Here we
want the
charge density, which is (minus) the normal derivative of the
electrostatic
potential due to $\sigma(x)$ just above the real axis between $a$ and
$b$,
given the tangential derivative for $a<x<b$ (which must cancel the
tangential
derivative of the given external electrostatic potential for the
charges to be
in equilibrium) and the fact that the normal derivative is zero
elsewhere just
above the real axis (and is zero at infinity), with $D$ being the
upper half
Euclidean plane.  In our special case, the solution is given by the
following
theorem \cite{Tri1,Tri2}:  Given the finite Hilbert transformation
	\begin{equation}
        f(x) = \frac{1}{\pi}\int _{-1}^{1}\frac{\phi(y)dy}{y-x},
	\end{equation}
the inverse Hilbert transformation is given by
	\begin{equation}
        \phi(x) = -\frac{1}{\pi}\int _{-1}^{1}
	\sqrt{\frac{1-y^2}{1-x^2}}\frac{f(y)dy}{y-x}
	+\frac{C}{\sqrt{1-x^2}},
	\end{equation}
where the principal parts are taken for both integrals, and where
	\begin{equation}
        C = \frac{1}{\pi}\int _{-1}^{1}\phi(y)dy
	\end{equation}
is an arbitrary constant.

Applying this theorem to our problem, singularities can be avoided at
$a$ and
$b$ if
	\begin{equation}
        a=2+w-2\sqrt{1+w},\;\;b=2+w+2\sqrt{1+w}.
	\end{equation}
Then the charge density (normalized density of eigenvalues of the
typical
$\rho_A$) is
	\begin{equation}
        \sigma(x)=\frac{\sqrt{-x^2+2(2+w)x-w^2}}{2\pi x}
	=\frac{\sqrt{(x-a)(b-x)}}{2\pi x}.
	\end{equation}
This gives, under our large $(m, n)$ approximation,
	\begin{eqnarray}
	S_{m,n} &\simeq &\ln n - \frac{m}{2n}\int_a^b\sigma(x)x\ln x
dx
	=\ln n - \frac{2}{\pi}\int_{-1}^{1}dy
y\sqrt{1-y^2}\ln(2+w+2\sqrt{1+w}y)
	\nonumber \\&=&\ln n - \frac{2}{\pi}
	\int_{0}^{2\pi}d\theta
\sin^2\theta\ln\sqrt{1+2r\cos\theta+r^2},
	\end{eqnarray}
with
	\begin{equation}
        y\equiv\frac{2x-a-b}{b-a}\equiv\cos\theta,\;\;
	r\equiv\sqrt{1+w}\equiv\sqrt{\frac{n}{m}}\geq 1.
	\end{equation}
The argument of the logarithm of the last integral is the distance
from a point
on the unit circle in the $(y,z)$ plane to a point at distance $r$
along the
negative real axis from the center.  Thus the integral can be viewed
as yet
another electrostatic potential in two dimensions, at $y=-r$ from a
$\sin^2\theta$ charge distribution around the unit circle, which is a
monopole
plus a quadrupole, and this works out to give
	\begin{equation}
	S_{m,n} \simeq \ln n-2\ln r - \frac{1}{2r^2} = \ln m -
\frac{m}{2n},
	\end{equation}
which is Eq. (10).

Thus we see that when the dimensions $m$ and $n$ of both subsystems
$A$ and $B$
are large, and when the joint system is in a random pure state, the
smaller
subsystem $A$ (with dimension $m$) typically has nearly maximal
entropy $\ln
m$.  The average deviation or information in the smaller subsystem is
	\begin{equation}
	I_{m,n} = \frac{m}{2n} + O(\frac{1}{mn}),
	\end{equation}
and is always less than one half of a natural logarithmic unit.  That
is, for a typical pure quantum state of a large system, the smaller
subsystem
is very nearly maximally mixed, showing little signs that the total
system is
pure.

Another way of putting it is to say that if the subsystems $A$ and
$B$ were
broken up into tiny sub-subsystems, which typically would each be
very nearly
maximally mixed, there would be virtually no information is the
sub-subsystems
considered separately.  For quantum information, the whole system
contains more
information than the sum of the information in the separate parts,
and in this
case almost all the information giving the precise pure state of the
entire
system, $\ln m+\ln n$ units, is in the correlations of the
sub-subsystems.  The
result above shows that for a typical pure state of the entire
system, very
little of the information, roughly $m/(2n)$ unit, is in the
correlations within
the smaller subsystem $A$ itself, roughly $\ln n -\ln m +m/(2n)$
units is in
the
correlations within the larger subsystem $B$ itself, and the
remaining roughly
$2\ln m -m/n$ units of information are in the correlations beween the
larger
and
smaller subsystems.

{\bf Acknowledgments}:  The hospitality of the Aspen Center for
Physics in
Colorado and of Kip Thorne and Carolee Winstein at their home in
Pasadena,
California, where the major part of this research was accomplished,
is
gratefully acknowledged, as is the California Institute of
Technology, the
University of California at Santa Barbara, and the Journ\'{e}es
Relativistes
'93 at the Universit\'{e} Libre de Bruxelles, where this work was
reported.
Thanks are also due Seth Lloyd for informing me of his paper with the
late
Heinz Pagels and clarifying what the differentials should be for
their
probability distribution.  At the April Brussels meeting, after my
calculations
were completed, Esteban Calzetta kindly informed me of  the earlier
work on
this subject by Elihu Lubkin, who then graciously sent me his more
recent paper
with Thelma Lubkin and provided useful discussions and corrections by
e-mail.
Financial support was provided in part by the
Natural Sciences and Engineering Research Council of Canada.

\baselineskip 19   pt
\pagebreak

\end{document}